\documentclass[12pt]{iopart}
\usepackage{amsmath,graphicx,amssymb}
\begin{document}
%%%%% NEWCOMMANDS %%%%%%%%%%%%%%%%%%%%%%%%%%%%%%%%%%%%%%%%%%%%%%%%%%%%%
\newcommand{\bmsigma}{\boldsymbol \sigma}
\newcommand{\bM}{\boldsymbol M}
\newcommand{\bV}{\boldsymbol V}
\newcommand{\media}[1]{\langle #1 \rangle}
\newcommand{\mbfrac}{\mbox{$ \frac12 $}}
\newcommand{\ket}[1]{\vert #1 \rangle}
%%%%%%%%%%%%%%%%%%%%%%%%%%%%%%%%%%%%%%%%%%%%%%%%%%%%%%%%%%%%%
\title{Characterization of bipartite states using a single homodyne 
detector}
\author{Virginia D'Auria, Alberto Porzio, and Salvatore Solimeno}
\address{CNR-INFM and Dipartimento di Scienze Fisiche Universit\`a 
`Federico II' Napoli, Italy}
\author{Stefano Olivares and Matteo G.A. Paris}
\address{Dipartimento di Fisica dell'Universit\`{a} di Milano, Italia.} 
\date{}
\begin{abstract}
We suggest a scheme to reconstruct the covariance matrix of a 
two-mode state using a single homodyne detector plus a polarizing 
beam splitter and a polarization rotator.  It can be 
used to fully characterize bipartite Gaussian states and to extract
relevant informations on generic states.
\end{abstract}
%%%%%%%%%%%%%%%%%%%%%%%%%%%%%%%%%%%%%%%%%%%%%%%%%%%%%%%%%%%%%
\section{Introduction}
Bipartite (entangled) states of two modes of the radiation
field are the basic tool of quantum information processing
with continuous variables \cite{SamRMP,eis,nap}. \par
Bipartite states can be produced by different schemes, mostly 
based on parametric processes in active nonlinear optical media. 
Generation schemes are either Hamiltonian two-mode processes, 
like parametric downconversion \cite{down} or mixing of squeezed 
states \cite{mixsq}, or conditional schemes based on the generation 
of multipartite  states followed by conditional measurements 
\cite{ale}. \par 
Besides mean values of the field operators, the most relevant 
quantity needed to characterize a bipartite state is its 
covariance matrix. For Gaussian states, a class 
that encompasses most of the states actually realized in quantum 
optical labs, the first two moments fully characterize the quantum 
state \cite{sim1,marians}. 
Once the covariance matrix is known then the entanglement 
of the state can be evaluated and, in turn, the performances
of the state itself in serving as a support for quantum
information protocols like teleportation or dense coding.
\par
Entanglement is generally corrupted by the interaction with 
the environment. Therefore, entangled states that are available 
for experiments are usually mixed states, and it becomes crucial 
to establish whether or not entanglement has survived the 
environmental noise. As a consequence, besides being of fundamental 
interest, a simple characterization technique for bipartite 
states is needed for experimentally check the accessible entanglement
in a noisy channel \cite{kim,praz,oli,twom,rec}, as well as the corresponding 
state purity and nonclassicality \cite{sal1,sal2}.
\par
In this paper we suggest a scheme to measure the first two moments 
of a bipartite state using repeated measurements of single-mode
quadratures made with a single homodyne detector. This is an improvement
compared to the scheme of Ref. \cite{rec}, where two homodyne detectors
have been employed.
The scheme involves fourteen quadratures pertaining to five different field
modes. It can be used to fully characterize bipartite Gaussian 
states or to extract relevant informations on a generic state.
\par
In the next Section we introduce the notation and describe how to obtain
the mean values and the covariance matrix starting from the statistics
of suitably chosen field quadratures. In Section \ref{s:exp} a possible 
experimental realization is described in details. Section \ref{s:out}
closes the paper with some concluding remarks.
%%%%
\section{Bipartite Gaussian states and reconstruction of the covariance
matrix}\label{s:the}
Our scheme is aimed to reconstruct the first two moments 
of a bipartite states. This represents a relevant piece of information
on any quantum state of two modes and provide the full characterization
of the quantum state in the case of Gaussian signals. 
Gaussian states,{\em i.e.} states with a Gaussian characteristic
function, are at the heart of quantum information processing
with continuous variables. The basic reason is that the vacuum state 
of quantum electrodynamics is itself a Gaussian state. This
observation, in combination with the fact that the quantum evolutions
achievable with current technology are described by Hamiltonian
operators at most bilinear in the quantum fields, accounts for the
fact that the states commonly produced in laboratories are Gaussian.
In fact, bilinear evolutions preserve the Gaussian character of the 
vacuum state \cite{sim2}. Furthermore, recall that
the operation of tracing out a mode from a multipartite Gaussian state
preserves the Gaussian character too, and the same observation 
is valid when the evolution of a state in a standard  noisy channel is 
considered. \par
We denote the two modes under investigation by $a$ and $b$. 
In the following we assume that  
$a$ and $b$ have equal frequencies and different polarizations.
The Cartesian operators $q_k$ and $p_k$, $k=a,b$ can be expressed in
terms of the mode operators as follows
\begin{eqnarray}
q_a = \frac{1}{\sqrt{2}}(a^\dag + a)\,, \qquad 
p_a = \frac{i}{\sqrt{2}}(a^\dag - a)\,,
\end{eqnarray}
and analogously for $q_b$ and $p_b$.
The covariance matrix of a two-mode state is a real symmetric positive 
matrix defined as follows
\begin{align}
\bmsigma =
\left(
\begin{array}{cccc}
\Delta q_a^2 & \Delta q_a p_a & \Delta q_a q_b & \Delta q_a p_b \\[1ex]
\Delta p_a q_a & \Delta p_a^2 & \Delta p_a q_b & \Delta p_a p_b \\[1ex]
\Delta q_b q_a & \Delta q_b p_a & \Delta q_b^2 & \Delta q_b p_b \\[1ex]
\Delta p_b q_a & \Delta p_b p_a & \Delta p_b q_b & \Delta p_b^2
\end{array}
\right)\,,
\end{align}
where $\Delta X^2 = \langle X^2 \rangle - \langle X \rangle^2$ and 
$\Delta XY = \frac12 \langle [X, Y]_{+}  \rangle - \langle X \rangle
\langle Y \rangle$ denote the variance of the observable $X$ and 
the and mutual correlations between the observables $X$ and $Y$
respectively. $[X, Y]_{+}=XY + YX$ denotes the anticommutator 
between the operators $X$ and $Y$.
Throughout the paper $\langle X \rangle$ will denote 
the ensemble average  $\langle X \rangle=\hbox{Tr}\left[R\: X\right]$,
$R$ being the density matrix describing the two-mode state.  
The characteristic function of a quantum state $R$ is defined as 
the expectation values $\chi(\lambda_1,\lambda_2)=\langle D(\lambda_1)
\otimes D(\lambda_2)\rangle$ where $\lambda_j\in {\mathbb C}$, $j=1,2$
and $D(\lambda)=\exp\left\{\lambda a^\dag - \lambda^* a\right\}$
is the displacement operator. The most general bipartite Gaussian state 
corresponds to a characteristic function of the form
\begin{equation}
\chi({\boldsymbol\lambda}) = \exp\left\{ -\mbox{$\frac12$} {\boldsymbol
\lambda}^T \bmsigma {\boldsymbol \lambda} - i {\boldsymbol \lambda}^T
{\boldsymbol X}\right\}\,,
\end{equation}
where ${\boldsymbol \lambda} = (\hbox{Re }[\lambda_1], \hbox{Im}[\lambda_1],
\hbox{Re }[\lambda_2], \hbox{Im}[\lambda_2])^T$ and $(\cdots)^T$ 
denotes transposition. The vector ${\boldsymbol
X}=(\media{q_a},\media{p_a},\media{q_b},\media{q_b})^T$ contains the 
mean value of the Cartesian mode operators. The characteristic function 
fully specify a quantum state, {\em i.e.} any expectation value
may be obtained as a phase space integral. Since for a Gaussian state
the first two moments specify the characteristic function, their
knowledge fully characterize a bipartite Gaussian state.
%%%%%%%%%%%%%%%%%
\subsection{Covariance matrix from quadrature measurement}
For the sake of simplicity, we rewrite the covariance matrix as 
follows:
\begin{equation}
\bmsigma = {\boldsymbol V}- {\boldsymbol M}
\end{equation}
where the variance $\bV$ and the mean $\bM$ matrices may 
be written as 
\begin{equation}\label{0:CVM}
\bV=
\left(
\begin{array}{c c c c }
\media{q_a^2} & \frac12 \media{[p_a, q_a]_{+}} &
\media{q_a q_b} & \media{q_a p_b}\\[1ex]
 \frac12 \media{[p_a, q_a]_{+}} & \media{p_a^2} &
\media{p_a q_b} & \media{p_a p_b} \\[1ex]
\media{q_b q_a} & \media{q_b q_a} &
\media{q_b^2} & \frac12 \media{[q_b, p_b]_{+}} \\[1ex]
\media{p_b q_a} & \media{p_b p_a} &
\frac12 \media{[p_b, q_b]_{+}} & \media{p_b^2}
\end{array}
\right)\,,
\end{equation}
and
\begin{equation}\label{M:CVM}
\bM=
\left(
\begin{array}{c c c c }
\media{q_a}^2 & \media{p_a} \media{q_a} &
\media{q_a} \media{q_b} & \media{q_a} \media{p_b}\\[1ex]
\media{p_a} \media{q_a} & \media{p_a}^2 &
\media{p_a} \media{q_b} & \media{p_a} \media{p_b} \\[1ex]
\media{q_b} \media{q_a} & \media{q_b} \media{q_a} &
\media{q_b}^2 & \media{q_b} \media{p_b} \\[1ex]
\media{p_b} \media{q_a} & \media{p_b} \media{p_a} &
\media{p_b} \media{q_b} & \media{p_b}^2
\end{array}
\right)\,.
\end{equation}
Once defined the quadrature operator of the mode $k$, namely
\begin{equation}
x_{k,\phi} = \frac{k^\dag\,e^{i\phi} + k\,e^{-i\phi}}{\sqrt{2}}\,,
\end{equation}
we use the following conventions:
\begin{subequations}
\begin{align}
&x_k \equiv x_{k,0}\,,\quad
y_k \equiv x_{k,\pi/2}\,, \\
&z_k \equiv x_{k,\pi/4}\,,\quad
t_k \equiv x_{k,-\pi/4}\,.
\end{align}
\end{subequations}
The matrix $\bM$ only contains the first moments 
and can be reconstructed by measuring the four 
quadratures $x_k$ and $y_k$, $k=a,b$. We have 
\begin{equation}
\media{q_k} = \media{x_k}\,, \quad \media{p_k} = 
\media{y_k}\,.
\end{equation}
In order to reconstruct the variance matrix $\bV$ more quadratures are 
needed. Let us introduce  the modes 
\begin{equation}\label{modes}
a,\quad b, \quad c = \frac{a+b}{\sqrt{2}}, \quad d = \frac{a-b}{\sqrt{2}},
\quad e = \frac{ia+b}{\sqrt{2}}, \quad f = \frac{ia - b}{\sqrt{2}}\,.
\end{equation}
If $a$ and $b$ correspond to vertical and horizontal polarizations, 
then $c$ and $d$ are rotated polarization modes at $\pm \pi/4$, whereas
$e$ and $f$ correspond to left- and right-handed circular polarizations.
After tedious but straightforward calculations, we have:
\begin{align}
\bV = \frac12 \left(
\begin{array}{c c c c }
2\media{x_a^2} & \media{z_a^2} -\media{t_a^2} & \media{x_c^2} - \media{x_d^2} & 
\media{y_e^2} - \media{y_f^2} \\[1ex] 
\media{z_a^2} -\media{t_a^2} & 2\media{y_a^2} & \media{x_f^2} 
- \media{x_e^2} & \media{y_c^2} - \media{y_d^2} \\[1ex]
\media{x_c^2} - \media{x_d^2} & \media{x_f^2} - \media{x_e^2} & 2\media{x_b^2} & 
\media{z_b^2} - \media{t_b^2} \\[1ex]
\media{y_e^2} - \media{y_f^2} & \media{y_c^2} - \media{y_d^2} & 
\media{z_b^2} - \media{t_b^2} & 2\media{y_b^2}
\end{array}
\right)\,.
\end{align}
Furthermore, since
\begin{subequations}
\begin{align} \label{qapb-paqb}
&\bV_{14} = \bV_{41} = \mbfrac \left(\media{y_e^2} - \media{y_f^2}\right)
= \media{y_e^2} - \mbfrac \left(\media{x_a^2} + \media{y_b^2}\right)\,,\\
&\bV_{23} = \bV_{32} =\mbfrac \left(\media{x_f^2} - \media{x_e^2}\right)
= \mbfrac \left(\media{x_b^2} + \media{y_a^2}\right) - \media{x_e^2}\,, 
\end{align}
\end{subequations}
the measurement of the quadratures pertaining to mode $f$ is not
essential. Overall, in our scheme, the reconstruction of the covariance matrix 
requires the measurement of at least fourteen quadratures, {\em e.g.} the
following ones (of course measuring also the $f$-quadratures, being 
additional independent measurements, would improve the accuracy of the
reconstruction) 
\begin{align}
\begin{array}{llll}
x_a, &y_a, &z_a, &t_a,\\
x_b, &y_b, &z_b, &t_b,\\
x_c, &y_c, &x_d, &y_d,\\
x_e, &y_e; & &
\end{array}
\nonumber\:.
\end{align}
Notice that the number of parameters needed to characterize a bipartite 
Gaussian state is also equal to fourteen.
%%%%%%%%%%%%%%%%%%%%%%%%%%%%%%%%%%%%%%%%%%%%%%%%%%%%%%%%%%%%%%%%%%
\section{Experimental implementations}\label{s:exp}
In Section \ref{s:the} we have proved that it is possible to fully 
reconstruct the covariance matrix $\bmsigma$ by measuring fourteen 
different quadratures of five field modes obtained as linear combination 
of the initial pair. Here we consider an implementation based on the 
bright continuous-wave beams generated by a seeded degenerate optical 
parametric amplifier (DOPA) below threshold based on a type--II 
nonlinear crystal \cite{peng}. The two collinear beams ($a$ and $b$) 
exiting the DOPA are orthogonally polarized and excited in a
continuous variable bipartite entangled state. In the following we
assume $a$ as vertically polarized and $b$ as horizontally polarized.
%%%%%%%
\begin{figure}[h]
\setlength{\unitlength}{1mm}
\begin{center}
\includegraphics[width=65\unitlength]{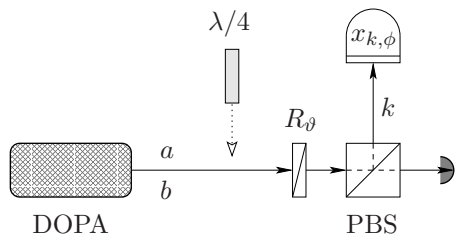}
\end{center}
\vspace{-.5cm}
\caption{Scheme of a possible apparatus to measure the covariance
matrix of the bipartite (entangled) state generated by a DOPA. The two modes, 
$a$ (vertical polarization) and $b$ (horizontal polarization), pass
through a (removable) $\lambda/4$ wave-plate and a rotator of polarization
$R_{\vartheta}$; finally, a PBS reflects the vertically polarized component
of its input toward a homodyne detector, which measures the $x_{k,\phi}$
quadrature. See text for details.} \label{f:S}
\end{figure}
%%%%%%%
\par
Since the mode $f$ is not necessary to reconstruct the covariance matrix,
we do not consider its selection, focusing our attention on modes $a$,
$b$, $c$, $d$, and $e$. The mode under scrutiny is selected by inserting
suitable components on the optical path of fields $a$ and $b$, before 
the homodyne detector. 
Modes $a$, $b$, $c$, and $d$ are obtained by means of a rotator of 
polarization $R_{\vartheta}$ (namely a $\lambda /2$ wave-plate) and 
a polarizing beam splitter (PBS), which reflects toward the detector 
the vertically polarized component of the impinging beam. 
The action of the rotator $R_{\vartheta}$ on the basis 
$\{\ket{V},\ket{H}\}$ is given by
\begin{subequations}
\begin{align}
R_{\vartheta} \ket{V} &= \cos\vartheta\,\ket{V} -
\sin\vartheta\,\ket{H}\,,\\
R_{\vartheta} \ket{H} &= \sin\vartheta\,\ket{V} +
\cos\vartheta\,\ket{H}\,.
\end{align}
\end{subequations}
In order to select mode $e$ a $\lambda /4$ wave-plate should be inserted just 
before the rotator $R_{\vartheta}$ (see Fig.~\ref{f:S}). The $\lambda/4$ wave-plate 
produces a $\pi/2$ shift between horizontal and vertical polarization
components, thus turning the polarization from linear into circular. 
\par
Table \ref{t:settings:S} summarizes the settings needed to select the five modes.  
Overall, the vertically polarized mode $k$ arriving at the detector 
can be expressed in terms of the initial modes as follows
\begin{equation}
k = \exp\{i\varphi\} \cos\vartheta\, a + \sin\vartheta\, b\,,
\end{equation}
where $\varphi = \pi/2$ when the $\lambda/4$ wave-plate is inserted,
$\varphi = 0$ otherwise. 
%%%%%%%%%%%%%%%%%%%%%%%%%%%%%%%%%%%%%%%%%%%%%%%%%%%%%%%%%%%%%%%%%%%
\begin{table}[h]
\begin{center}
\begin{tabular}{|c|c|c|c|c|c}
\hline
Mode & $\lambda/4$ & $R_{\vartheta}$ \\
\hline
$a$ & no & $0$ \\
\hline
$b$ & no & $+\pi/2$ \\
\hline
$c$ & no & $+\pi/4$ \\
\hline
$d$ & no & $-\pi/4$ \\
\hline
$e$ & yes & $+\pi/4$ \\
\hline
\end{tabular}
\end{center}
\caption{Setting to select the different modes $k$. The table refers to the 
elements depicted in Fig.~\ref{f:S}. The mode $a$ is assumed to be
vertically polarized and the mode $b$ horizontally polarized.}
\label{t:settings:S}
\end{table}
\par
Once the mode $k$ has been selected, a homodyne detector is used to
measure the generic quadrature $x_{k,\phi}$. Homodyne relies on the
controlled interference between the quantum beam (signal) to be analyzed and
a strong ``classical'' local oscillator (LO) beam of phase $\phi$. Indeed,
to access $x_{k,\phi}$ one have to suitably tune the phase $\phi$. The
optimization of the efficiency is provided by matching the LO mode to
the mode $k$. 
The mode matching requires precise control of the LO frequency, spatial and
polarization properties.
Remarkably, the detected mode is always vertically
polarized, thus avoiding any need of tuning the LO polarization.
%%%%%%%%%%%%%%%%%%%%%%%%%%%%%%%%%%%%%%%%%%%%%%%%%%%%%%%%%%%%%%%%%%
\section{Conclusions}\label{s:out}
A simple scheme has been suggested to reconstruct the covariance matrix 
of two-mode states of light using a single homodyne detector plus a polarizing 
beam splitter and a polarization rotator. Our scheme requires
the local measurements of 14 different quadratures pertaining to five
field mode. It can be used to fully characterize bipartite Gaussian
states and to extract relevant informations on generic states.
Finally, we notice that an efficient source of polarization squeezing 
has been recently realized \cite{ulr}, which might be considered as a 
preliminary stage for the experimental realization of the present
characterization scheme.
%%%%%%%%%%%%%%%%%%%%%%%%%%%%%%%%%%%%%%%%%%%%%%%%%%%%%%%%%%%%%%%%%%

%%%%%%%%%%%%%%%%%%%%%%%%%%%%%%%%%%%%%%%%%%%%%%%%%%%%%%%%%%%%%
\end{document}